\documentclass[twocolumn,showpacs,aps,prd,nobibnotes,nofootinbib,floatfix]{revtex4-2}

\usepackage{amsmath}
\usepackage{graphicx,subfigure}
\usepackage[usenames]{color}
\usepackage[normalem]{ulem}
\usepackage{textcomp}
\usepackage{gensymb}
\usepackage{xspace}
\usepackage{verbatim}

\usepackage{soul}

\usepackage[colorlinks=true]{hyperref}
\hypersetup{
  citecolor  =red,
  urlcolor=blue
}
\usepackage{orcidlink}
\newcommand{\rh}{r_{\text{h}}}

\begin{document}

\title{ Holographic thermodynamics of a five-dimensional neutral Gauss-Bonnet AdS black hole}

\author{Si-Jiang Yang\orcidlink{0000-0002-8179-9365}$^a$$^b$$^c$}

\author{Md Sabir Ali\orcidlink{0000-0001-6670-7955}$^a$$^b$$^c$$^d$}

\author{Shao-Wen Wei\orcidlink{0000-0003-0731-0610}$^a$$^b$$^c$}

\author{Yu-Xiao Liu\orcidlink{0000-0002-4117-4176}$^a$$^b$$^c$}%
\email{liuyx@lzu.edu.cn, corresponding author}
\affiliation{$^{a}$Key Laboratory of Quantum Theory and Applications of MoE, Lanzhou Center for Theoretical Physics, Lanzhou University, Lanzhou 730000, China\\
$^{b}$Key Laboratory of Theoretical Physics of Gansu Province, Institute of Theoretical Physics $\&$ Research Center of Gravitation, Lanzhou University, Lanzhou 730000, China \\
$^{c}$School of Physical Science and Technology, Lanzhou University, Lanzhou 730000, China \\
$^{d}$Department of Physics, Mahishadal Raj College, West Bengal 721628, India}
\date{\today}

\begin{abstract}
Motivated by the recent progress on the holographic dual of the extended thermodynamics for black holes in anti-de Sitter (AdS) space, we investigate the holographic thermodynamics for the five-dimensional neutral Gauss-Bonnet AdS black hole in the context of the anti-de Sitter/conformal field theory (AdS/CFT) correspondence. Through the extended bulk thermodynamics for the five-dimensional Gauss-Bonnet AdS black hole, we derive the first law of the CFT thermodynamics which is obtained by directly translating the arbitrary conformal factors in the dual CFT.
In addition to the newly defined chemical potential $\mu$ conjugating to the central charge $C$, we obtain other pairs of thermodynamics for the CFT, such as the temperature $\tilde{T}$ and the entropy $S$, the Gauss-Bonnet coupling constant $\tilde{\alpha}$ and its conjugate variable $ \tilde{\mathcal{A}}$, the pressure $\mathcal{P}$ and its conjugate volume $\mathcal{V}$. In the fixed $C$, $\mathcal{V}$ and $\tilde{\alpha}$ canonical ensemble, we obtain the canonical description of the CFT thermodynamics and observe the standard swallowtail behavior in the Helmholtz free energy vs the temperature plot. The self-intersection point of the Helmholtz free energy indicates the phase transition between the high and low entropy states of the CFT. By using Maxwell's equal area law, we get the critical point and coexistence curve for the high and low entropy phases of the CFT. Besides, we get the critical exponents for the CFT, and find that the critical point and critical exponents associated with the $\tilde{T}-S$ criticality of the CFT are the same as those of the five-dimensional Gauss-Bonnet AdS black hole.
\end{abstract}

\maketitle

\section{Introduction}\label{sec:intro}
Black hole thermodynamics in asymptotically anti-de Sitter (AdS) spacetime has advanced to the forefront of current research activities in theoretical high energy physics. One of the intriguing aspects of theoretical physics is duality which has revolutionized perspectives of physics at a fundamental level. Dualities relate two different facets of nature and thus they lead to an enhanced understanding of physics. An implication of these dualities is that, while the mathematical structures of the two theories may be identical, there may exist differences from a physical standpoint. One of the interesting dualities is the anti-de Sitter/conformal field theory (AdS/CFT) correspondence~\cite{Maldacena:1997re,Gubser:1998bc,Witten:1998qj} which connects the CFT on a flat spacetime to string theory. This is interesting because string theory is a theoretical framework for quantum gravity, whereas quantum field theory is not. 

Now, even though these two theories are different, the AdS/CFT correspondence assumes the existence of a correspondence between the two theories and this is one of the most attractive features of this duality, which pushes research for its deeper understanding and possible applications. A renowned example of a phase transition in black hole spacetimes is the radiation/black hole first-order phase transition of Hawking and Page, observed for Schwarzschild-AdS black holes immersed in a bath of radiation~\cite{HaPa83}. Such a phenomenon has a dual interpretation for a boundary quantum field theory via the AdS/CFT correspondence and is related to a confinement/deconfinement phase transition in the dual quark gluon plasma~\cite{Witt98,Witten:1998qj}. Further investigation of Hawking-Page phase transition for higher dimensional Schwarzshild-like AdS black hole in higher derivative gravity from the AdS/CFT correspondence perspective can be found in Refs.~\cite{Nojiri:2001aj,Cvetic:2001bk,Nojiri:2002qn}. For charged black hole in AdS spacetime one likewise observes a small/large black hole first-order phase transition reminiscent of the liquid/gas phase transition of the Van der Waals fluid~\cite{CEJM99a,CEJM99b}. But when the cosmological constant is interpreted as the thermodynamic pressure and the black hole mass is interpreted as the thermodynamic enthalpy~\cite{KaRT09,Dola11}, the phase behavior of the charged AdS black hole is exactly the same with that of the Van der Waals fluid and they have the same critical exponents~\cite{KuMa12}. This version of the black hole thermodynamics in the presence of negative cosmological constant is interpreted as the extended phase space thermodynamics or generically speaking, the black hole chemistry~\cite{KuMT17,Mann:2016trh}. The phase structures in the extended phase space is much richer, including reentrant  phase transition~\cite{Gunasekaran:2012dq,Frassino:2014pha}, superfluid-like phase transition~\cite{Hennigar:2016xwd}, triple points~\cite{Altamirano:2013uqa,Altamirano:2013ane} etc. Recent advancements have yielded numerous successful interpretations of the origins of the extended phase space thermodynamics. Notable among these are the extended Iyer-Wald formalism~\cite{XiTL24}, the coupling constant of gauge field perspective~\cite{HaTe24}. Additionally, a mechanism for the higher-dimensional origin of a dynamical cosmological constant was proposed~\cite{fpsv23}.

Despite significant advancements have been achieved in the field of thermodynamics for black holes in AdS spacetime, the holographic interpretation of black hole chemistry has remained an enigma~\cite{Mann24}. From the viewpoint of the AdS/CFT correspondence, the negative cosmological constant $\Lambda$ sets the asymptotic structure of the bulk spacetime and is related to the number of colours $N$ in the dual gauge field theory~\cite{Mald97,KaRo15}
\begin{equation}
    k\frac{L^{d-2}}{16\pi G}=N^2,
\end{equation}
where $L$ is the AdS radius and the numerical factor $k$ depends on the details of the particular holographic system. Hence, variation of the negative cosmological constant $\Lambda$ corresponds to changing the number of colors $N$ and hence varying the dual gauge field theories in the boundary~\cite{Johnson:2014yja,CKMV22,CoKM21}.

Recently there have been considerable progress in developing an exact dictionary between the black hole chemistry and the thermodynamics of the dual gauge field theory. Considering the viewpoint that Newton's constant is treated as a variable~\cite{CoKM21}, and extended to the more general CFT metric with boundary radius different from the AdS radius, Visser argued that holographic thermodynamics requires a chemical potential for color and obtained the thermodynamic Euler equation for high energy states of large-$N$ gauge theories~\cite{Viss22}. Systematic works have been done based on this viewpoint~\cite{Zheng:2024glr,Zhang:2023uay,Sadeghi:2023tuj,Ali:2023jox,Punia:2023ilo}. 
Building on Visser’s viewpoint that the radius of the CFT metric need not coincide with the AdS radius, and allowing the conformal factor to vary while keeping Newton’s constant fixed, thereby resulting in the boundary volume $\mathcal{V}$ and the central charge $C$ being completely independent,  Ahmed et al. constructed an exact dictionary between the thermodynamics of the charged AdS black hole and the dual conformal field theory~\cite{ACKMV23a}. Based on this perspective, a lot of work has been done for the holographic thermodynamics for AdS black holes, including the charged AdS black hole~\cite{Ladghami:2024wkv}, Kerr AdS black hole~\cite{GoJZ23,ACKMV23}, Kerr-Newman AdS black hole~\cite{Baruah:2024yzw}. Building on this perspective, we extend the study of the holographic dictionary to five-dimensional Gauss-Bonnet gravity, with a focus on the holographic thermodynamics and a comparison to the thermodynamics of the five-dimensional neutral Gauss-Bonnet AdS black hole. In our analysis, we treat Newton's constant as a fixed parameter, while allowing the conformal factor to vary.

From the perspective of the gauge/gravity correspondence, the thermodynamics of AdS black holes is completely equivalent to the thermodynamics of the dual gauge fields. Through the extended bulk thermodynamics for the five-dimensional Gauss-Bonnet AdS black hole, we derive the first law of the CFT thermodynamics which is obtained by directly translating the arbitrary conformal factors in the dual CFT and investigate the phase behavior and criticality of the CFT. We find that the free energy exhibits a characteristic swallowtail behavior and the critical points and critical exponents are the same as that of the five-dimensional Gauss-Bonnet AdS black hole.

The outline of the paper is as follows. In Sec.~\ref{second section}, we derive the AdS/CFT dictionary and the laws for the CFT residing in the boundary dual to the five-dimensional Gauss-Bonnet AdS black hole. In Sec.~\ref{third section}, we investigate the phase behavior and criticality of the CFT. Finally, in the last section we discuss the results and conclude them. 

\section{Holographic thermodynamics of five-dimensional Gauss-Bonnet AdS black hole}\label{second section}

The thermodynamic for black holes in AdS space-time has attracted a lot of attention and also raise perplexities. There are diverse perspectives on the thermodynamics of five-dimensional Gauss-Bonnet AdS black holes. From the viewpoint that the Gauss-Bonnet coupling constant is treated as an independent thermodynamic variable, the thermodynamics and phase transitions of Gauss-Bonnet AdS black holes have been systematically studied~\cite{Wei:2012ui,CCLY13,Wei:2014hba}. Adopting the perspective that Newton's constant is allowed to vary and introducing a central charge, Kumar et al. derived a first law of thermodynamics that includes contributions from both bulk and boundary variables, known as the mixed form of the first law~\cite{Kumar:2022afq}. By varying Newton's constant further and proposing the existence of two central charges: $A$-charge and $C$-charge, Fan et al. developed a new first law of black hole thermodynamics for Gauss-Bonnet AdS black holes~\cite{CuFa24}. 

In this section, we explore the extended thermodynamics of the five-dimensional neutral Gauss-Bonnet AdS black hole, based on the perspective that the cosmological constant and the Gauss-Bonnet coupling constant are treated as independent thermodynamic variables, while Newton's constant is fixed~\cite{Wei:2012ui,CCLY13}, we tentatively establish a connection between the extended thermodynamics of the five-dimensional neutral Gauss-Bonnet AdS black hole and the corresponding extended thermodynamics of dual CFT. By exploring the relationship between these two realms, we aim to shed light on the profound similarities and underlying principles that govern their thermodynamic properties.

\subsection{Extended bulk thermodynamics}

In this subsection, we briefly review the main points of the extended black hole thermodynamics for the five-dimensional neutral Gauss-Bonnet AdS black hole. A $d$-dimensional neutral Gauss-Bonnet AdS black hole is a solution for the gravitational field equation derived by the action~\cite{BoDe85}
\begin{equation}\label{Action}
  S=\int d^dx\sqrt{-g}\frac{1}{16\pi G_N}\left(\mathcal{R}-2\Lambda+\alpha_{\text{GB}}\mathcal{L_{\text{GB}}}\right) 
\end{equation}
with the Gauss-Bonnet term
\begin{equation}\label{GBterm}
\mathcal{L_{\text{GB}}}=\mathcal{R}_{\mu\nu\alpha\beta}\mathcal{R}^{\mu\nu\alpha\beta} -4\mathcal{R}_{\mu\nu}\mathcal{R}^{\mu\nu}+\mathcal{R}^2,
\end{equation}
where $G_N$ is the Newton constant, $\alpha_{\text{GB}}$ is the Gauss-Bonnet coupling constant, and the cosmological constant is given by
\begin{equation}
    \Lambda=-\frac{(d-1)(d-2)}{2L^2}
\end{equation}
with $L$ the AdS radius.

Within the context of the low energy effective action formalism in heterotic string theory, the Gauss–Bonnet term is the leading order quantum correction to gravity~\cite{GrSl87}. The Gauss–Bonnet coupling constant is related to the string scale and can be identified with the inverse of the string tension~\cite{ChDu02,Wei:2012ui,YWCYW20}. Hence, we only consider positive Gauss-Bonnet coupling constant in this paper.

The theory admits an exact neutral black hole solution in five dimensions. The metric for the five-dimensional neutral Gauss-Bonnet AdS black hole is~\cite{Cai02}
\begin{equation}\label{GBmetric}
  ds^2=-f(r)dt^2+f^{-1}(r)dr^2+r^2d\Omega_3^2,
\end{equation}
where $d\Omega_3^2$ is the metric on the three-dimensional unit sphere:
\begin{equation}\label{3sphere}
  d\Omega_3^2=d\theta^2+\sin^2\theta (d\phi^2+\sin^2\phi d\psi^2),
\end{equation}
and the metric function $f(r) $ is given by
\begin{equation}\label{metricF}
  f(r)=1+\frac{r^2}{2\alpha}\left(1-\sqrt{1+\frac{32\alpha G_{\text{N}}M}{3\pi r^4}+\frac{2\alpha \Lambda}{3}}\right),
\end{equation}
where $\alpha=2\alpha_{\text{GB}}$.

The solution is a steady spherical solution with the event horizon $r_{\text{h}}$ determined by $f(r_{\text{h}})=0$. From this equation, we can express the mass of the black hole as
\begin{equation}\label{Mfunction}
  M=\frac{\pi}{8G_{\text{N}}}\left(-\frac{1}{2}\Lambda r_{\text{h}}^4+3r_{\text{h}}^2+3\alpha   \right).
\end{equation}
From the extended black hole thermodynamic perspective, the mass of the black hole is interpreted as the enthalpy $H=M$ rather than the internal energy $U$ of the gravitational system.

By requiring the absence of a conical singularity at the event horizon in the Euclidean action, we can get the Hawking temperature of the black hole, which is given by
\begin{equation}\label{HawkTemp}
  T_{\text{H}}=\frac{f'(r_{\text{h}})}{4\pi}=\frac{3r_{\text{h}}-\Lambda r_{\text{h}}^3}{6\pi r_{\text{h}}^2+12\pi\alpha}.
\end{equation}
From the viewpoint of the extended black hole thermodynamics, the cosmological constant $\Lambda$ is interpreted as the thermodynamic pressure and the black hole mass is regarded as the thermodynamic enthalpy $H$~\cite{KaRT09}. The thermodynamic pressure is
\begin{equation}\label{PresureLambda}
  P=-\frac{\Lambda}{8\pi G_{\text{N}}}=\frac{3}{4\pi G_{\text{N}} }\frac{1}{L^2}.
\end{equation}
From the perspective of extended black hole thermodynamics, the thermodynamic volume can be defined either as a surface integral of the Killing potential~\cite{KaRT09} or as the Killing volume~\cite{Jacobson:2018ahi,ACKMV23}
\begin{equation}
    V=\int_{\Sigma_{\text{bh}}}\mid\mid \xi \mid\mid dV-\int_{\Sigma_{\text{AdS}}}\mid\mid \xi \mid\mid dV,
\end{equation}
where $\mid\mid \xi \mid\mid=\sqrt{-\xi \cdot \xi}$ is the norm of the event horizon generating Killing vector $\xi=\partial_t$. The thermodynamic volume for the five-dimensional Gauss-Bonnet AdS black hole is
\begin{equation}\label{VolumeBH}
  V=\frac{\pi^2 r_{\text{h}}^4}{2}.
\end{equation}
In the low-energy effective action of heterotic string theory, the Gauss-Bonnet coupling constant can be identified with the inverse of the string tension, making it natural to treat the Gauss-Bonnet constant as a thermodynamic variable. Furthermore, in order to obtain a consistent first law of black hole thermodynamics and the Smarr relation for the Gauss-Bonnet AdS black hole while the negative cosmological constant is regarded as the thermodynamic pressure and Newton's constant is treated as a fixed parameter, the Gauss-Bonnet coupling constant $\alpha$ must be treated as a thermodynamic variable. The corresponding conjugate thermodynamic variable is given by
\begin{equation}\label{Avariable}
  \mathcal{A}=\left(\frac{\partial M}{\partial \alpha}\right)_{S,P}=-\frac{\pi}{8G_{\text{N}}}\frac{-4\Lambda r_{\text{h}}^4+9r_{\text{h}}^2-6\alpha}{r_{\text{h}}^2+2\alpha}.
\end{equation}
The entropy of the black hole is
\begin{equation}\label{HBEntropy}
  S=\int_0^{r_h}\left(\frac{\partial M}{\partial r_h}\right)_{P,\alpha}dr =\frac{\pi^2 r_{\text{h}}}{2G_{\text{N}}}(r_{\text{h}}^2+6\alpha).
\end{equation}
The result is consistent with the Euclidean path integral approach, as presented in~\cite{Myers:1988ze,Haroon:2020vpr}, and is also in agreement with the Wald entropy formulation~\cite{Iyer:1994ys}.
Evidently, the presence of the Gauss-Bonnet correction to Einstein's theory of gravity leads to a notable revision of the Bekenstein-Hawking area law when considering the black hole's temperature as its surface gravity, as emphasized in Ref.~\cite{KuLi23}.

In the perspective of the fixed Newton's constant, the consistent first law of black hole thermodynamics and the Smarr relation for the five-dimensional Gauss-Bonnet AdS black hole are~\cite{CCLY13}
\begin{eqnarray}
  dM &=& TdS+VdP+ \mathcal{A}d\alpha, \label{bulk1stlow}\\
  2M &=& 3TS-2PV+2\mathcal{A} \alpha.\label{SmarrR}
\end{eqnarray}
Unlike the perspective where Newton's constant is treated as a variable, the first law and Smarr relation in this framework are independent of the boundary variables. Furthermore, treating the Gauss-Bonnet coupling constant as an independent thermodynamic variable is consistent with recent derivations using the extended Iyer-Ward formalism, where the coupling constant is interpreted as a conserved charge in black hole thermodynamics~\cite{HaTe24}.

The first law and the Smarr relation are the starting point of our derivation of the first law of boundary thermodynamics. Based on the first law of black hole thermodynamics and the Smarr relation, we proceed to derive the boundary thermodynamics in the next subsection. This derivation establishes a tentative relationship between the thermodynamic properties of the bulk system and its boundary counterpart, allowing for a comprehensive analysis of their interconnected behavior. The insights gained from this derivation might enhance our understanding of the overall thermodynamic framework.

\subsection{Extended boundary thermodynamics}\label{boundarythermo}

Recently, Ahmed et al. proposed an approach that treats the conformal factor as a freely varying parameter, to solve the problem of the fixed Newton's constant causing interdependence between the central charge and boundary volume. This perspective yields a holographic first law that corresponds to the extended thermodynamics for charged rotating AdS black holes~\cite{ACKMV23a}. In this subsection, we extend the methodology developed in~\cite{ACKMV23a} to the Gauss-Bonnet AdS black hole, deriving a holographic first law dual to Eq.~\eqref{bulk1stlow}.

From the perspective of the AdS/CFT correspondence, the dual CFT lives on the conformal boundary of the asymptotically AdS spacetime. According to the recent development of the holographic chemistry~\cite{Mann24}, the boundary metric of the CFT dual to the five-dimensional Gauss-Bonnet AdS spacetime is~\cite{Gubser:1998bc,KaRo15}:
\begin{equation}
	ds^2=\omega^2\left(-dt^2+L^2d\Omega_3^2 \right),
\end{equation}
where $\omega$ is a dimensionless conformal factor. The standard approach is to set $\omega=1$, in which case the boundary volume $\mathcal{V}$ is proportional to $L^{3}$. Consequently, variations in the cosmological constant imply variations in the boundary CFT volume $\mathcal{V}$, suggesting the presence of a pressure-volume work term $\mathcal{P}d\mathcal{V}$, in the first law of CFT thermodynamics. However, this leads to the conclusion that the central charge $C$ and the thermodynamic volume $\mathcal{V}$ are not truly independent~\cite{ACKMV23a}. As proposed in~\cite{ACKMV23a}, we instead choose the conformal factor $\omega=R/L$, which is allowed to vary freely~\cite{Viss22,ACKMV23a,ACKMV23}, where $R$ is the radius of the boundary. This makes the boundary volume and the central charge independent thermodynamic variables. The variation of the conformal factor reflects the conformal symmetry of the boundary theory. 

From the first law of black hole thermodynamics~\eqref{bulk1stlow} and the Smarr relation~\eqref{SmarrR} for the five-dimensional Gauss-Bonnet AdS black hole, we can express the first law in terms of the boundary thermodynamics
\begin{multline}
      \delta\left(\frac{M}{\omega}\right)=\left(\frac{T}{\omega}\right)\delta S+\frac{M-TS-\mathcal{A}\alpha}{\omega}\frac{\delta\left(L^3/G_{\text{N}} \right)}{\left(L^3/G_{\text{N}} \right)}\\
       +\frac{\sqrt{G_{\text{N}}}\mathcal{A}}{\omega L}\delta\left(\frac{\alpha L}{\sqrt{G_{\text{N}}}} \right)-\left(\frac{\tilde{E}}{3\mathcal{V}}\right)\delta \mathcal{V},
\end{multline}
where $\mathcal{V}=2\pi^2 R^3$ is the spatial volume of the boundary on which the CFT resides.
In the derivation of the first law in terms of the boundary thermodynamics, the Newtown constant $G_{\text{N}}$ is regarded as a fixed constant, but the conformal factor $\omega $ is free to vary.

In the gauge/gravity duality, the central charge of the CFT is related to the AdS radius $L$ and the gravitational coupling constant $G_{\text{N}}$ in the bulk theory. As the authors of Ref.~\cite{Kumar:2022afq} did for the mixed first law of black hole thermodynamics for Gauss-Bonnet AdS black hole, the central charge is chosen as
\begin{equation}
     C=\frac{\Omega_{d-2} L^{d-2}}{16\pi G_{\text{N}}},
\end{equation}
where $\Omega_{d-2}$ is the volume of a $(d-2)$-dimensional unit sphere. Since the precise coefficient of the central charge is irrelevant for our purpose and only the combination $dC/C$ appears in the first law of the CFT, the dimensional dependent coefficient is normalized such that it agrees with the coefficient $A$ of the Euler density in the trace anomaly~\cite{CKMV22,Myers:2010tj,Chen:2023pgs}.

To get a consistent formulation of the first law of the CFT, we employ the following holographic dictionary:
\begin{equation}
    \begin{split}
         \tilde{E}&=\frac{M}{\omega}, \qquad \qquad \tilde{T}=\frac{T}{\omega}, \qquad  S=S, \\ 
    \tilde{\mathcal{A}}&=\frac{\sqrt{G_{\text{N}}}\mathcal{A}}{\omega L},
    \qquad \tilde{\mathcal{\alpha}}=\frac{\alpha L}{\sqrt{G_{\text{N}}}},\quad \mathcal{P}=\frac{\tilde{E}}{3\mathcal{V}}, \\
      C&=\frac{\pi L^3}{8G_{\text{N}}},  \qquad    \mu=\frac{1}{C}\left(\tilde{E}- \tilde{T} S-\tilde{\mathcal{A}} \tilde{\mathcal{\alpha}} \right),
    \end{split}
\end{equation}
the extended first law of thermodynamics for the five-dimensional Gauss-Bonnet AdS black hole is dual to the following thermodynamic first law in the CFT:
\begin{equation}
    \delta \tilde{E}=\tilde{T}\delta S+\tilde{\mathcal{A}}\delta \tilde{\alpha}+\mu\delta C-\mathcal{P}\delta \mathcal{V}, \label{1stlaw}
\end{equation}
where $\mathcal{P}$ is the field theory pressure, and $\mu$ is the chemical potential associated with the central charge $C$.  The generalized Smarr relation~\eqref{SmarrR} for the AdS black hole is dual to an Euler equation in a large~$N$ gauge field theory, which takes the following form:
\begin{align}
 \tilde{E}=\tilde{T} S+\tilde{\mathcal{A}} \tilde{\mathcal{\alpha}}+\mu C.\label{EularEq}
\end{align}
One noticeable point is that the Euler equation does not include a $\mathcal{PV}$ term.

From the above AdS/CFT dictionary, the central charge $C$ in the extended boundary thermodynamics is dual to the pressure $P$ of the extended bulk thermodynamics. Hence it is natural to take the viewpoint that the conjugate thermodynamics of the central charge---the chemical potential $\mu$ in the extended boundary thermodynamic is dual to the thermodynamical volume $V$ of the extended black hole thermodynamics. 

We define the dimensionless parameters $x$ and $y$ as
\begin{equation}\label{DimensionlessP}
  x=\frac{\rh}{L}, \qquad y=\frac{\alpha}{L^2}.
\end{equation}
Then, in terms of the parameters $x$ and $y$, the CFT thermodynamic quantities are given by
\begin{itemize}
  \item energy:
  \begin{equation}\label{CFTE}
  \tilde{E}  =\frac{3C}{R}(x^4+x^2+y),
    \end{equation}
  \item temperature:
  \begin{equation}\label{CFTT}
    \tilde{T}=\frac{1}{2\pi R}\frac{2x^3+x}{x^2+2y},
  \end{equation}
  \item entropy:
  \begin{equation}\label{CFTS}
    S  =4\pi Cx(x^2+6y),
  \end{equation}
  \item pressure:
  \begin{equation}\label{CFTP}
    \mathcal{P} =\frac{C}{2\pi^2R^4}(x^4+x^2+y),
  \end{equation}
  \item chemical potential:
  \begin{equation}\label{CFTchemicalPotential}
 \mu=\frac{1}{R}\frac{x^2[-x^2(x^2-1)+6y(x^2+1)]}{x^2+2y},
\end{equation}
  \item the Gauss coupling constant $\tilde{\alpha}$ and its conjugate CFT thermodynamic quantity $\tilde{\mathcal{A}}$:
  \begin{equation}\label{CFTA}
  \begin{split}
    \tilde{\alpha}&=\sqrt{\frac{8CL^3}{\pi}}y, \\
    \tilde{\mathcal{A}} &=-\frac{1}{R}\sqrt{\frac{\pi C}{8L^3}}\frac{24x^4+9x^2-6y}{x^2+2y}.
      \end{split}
  \end{equation}
\end{itemize}


Having the thermodynamics for the CFT dual to the five-dimensional Gauss-Bonnet AdS black hole, we delve into the exploration of the phase transition  and critical phenomena for the CFT.

\section{Thermodynamic ensembles in the dual CFT}\label{third section}

The extended thermodynamics and phase transition for CFT states that are dual to the singly-spinning asymptotically AdS black hole were investigated by Ahmed et al., and various phase transitions and critical behavior in the CFT were uncovered~\cite{ACKMV23}. Gong et al. found that the critical exponents for the CFT states in the vicinity of the critical point are the same with that of the Kerr-Newman AdS black hole~\cite{GoJZ23}. In this section, we investigate the phase behavior and criticality exhibited by the canonical ensemble within the CFT
framework.

\subsection{Canonical ensemble: swallowtail behavior and critical point}

The canonical ensemble for five-dimensional Gauss-Bonnet AdS black hole has been well study in the framework of extended black hole thermodynamics, i.e. black hole chemistry~\cite{CCLY13,WeLi20}. In these studies it was found that the five-dimensional Gauss-Bonnet AdS black hole displays the swallowtail behavior, which indicates the small/large black hole phase transition reminiscent of the Van der Waals like phase transition. Besides, it was shown that the critical point for the five-dimensional Gauss-Bonnet AdS black hole phase transition is~\cite{CCLY13}
\begin{equation}\label{BHcritical}
    r_{\text{hc}}=\sqrt{6\alpha}, \qquad P_c=\frac{1}{48\pi\alpha},  \qquad T_c=\frac{1}{2\sqrt{6\alpha}\pi}.
\end{equation}
In this subsection, we investigate the phase behavior of the CFT states dual to the five-dimensional neutral Gauss-Bonnet AdS spacetime.

From the previous subsection for the study of the boundary thermodynamics, we can get the relevant Helmholtz free energy in the fixed $(C, \mathcal{V}, \tilde{\alpha})$ canonical ensemble:
\begin{equation}\label{CaF}
    F=\tilde{E}-\tilde{T}S=-\frac{C}{R}\frac{x^6+18x^4y-x^4+3x^2y-6y^2}{x^2+2y}.
\end{equation}

\begin{figure}
  \begin{center}
\subfigure[]{\includegraphics[width=3in]{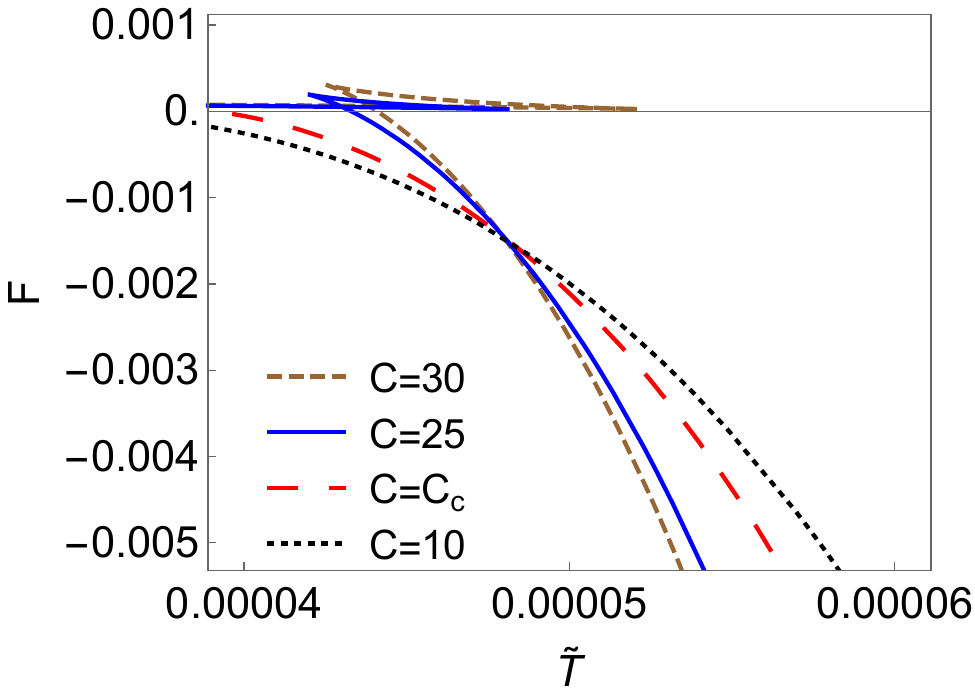}\label{Fig:swollowtail0}}
\subfigure[]{\includegraphics[width=3in]{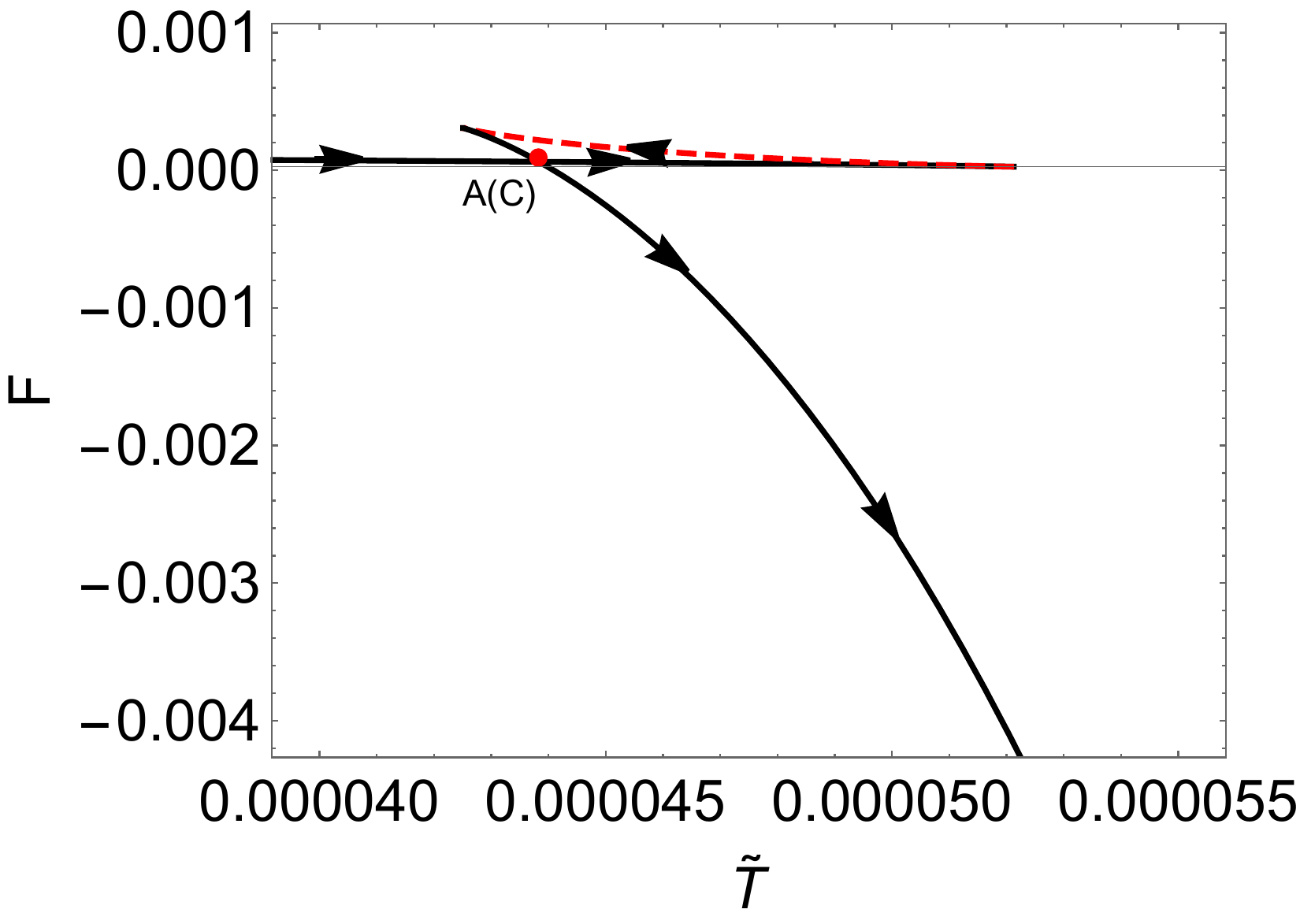}\label{Fig:FTdiagram1}}
    \caption{The $F-\tilde{T}$ diagram of the fixed $(C, \mathcal{V}, \tilde{\alpha})$ ensemble for the boundary CFT. We have set the Newtown's constant $G_{\text{N}}=1$, the boundary radius $R = 10,000$ and the coupling constant $\tilde{\alpha}=1$ in both figures.  (a) First-order phase transition occurs for $C>C_{\text{c}}=9\pi/2$ (the brown dashed and blue curves) as the ``swallowtail'' behavior indicates. There is a second-order phase transition for $C=C_{\text{c}}$ (the red large-dashed curve). No phase transition occurs for $C<C_{\text{c}}$ (the black dotted curve).  (b)  The entropy increases along the arrows attached to the curve. We have chosen the central charge  $C=30$. The heat capacities for the small and large entropy branches are both positive, hence they are thermodynamically stable; while the heat capacity of the intermediate entropy branch denoted by the red dashed line is negative, hence it is thermodynamically unstable.}\label{swallow}
  \end{center}
\end{figure}

To illustrate the phase behavior of the CFT states, we draw the diagram for the Helmholtz free energy $F$ vs the temperature $T$ in Fig.~\ref{swallow}. 
Though we have chosen the boundary radius $R=10^4$ and the thermodynamic variable of the CFT $\tilde{\alpha}=1$ in the figures, the scale of the boundary radius $R$ and $\tilde{\alpha}$ do not affect the qualitative thermodynamic behavior of the system. The figures show the characteristic swallowtail behavior which is similar to the liquid/gas phase transition for the Van der Waals fluid. The brown dashed and blue curves in Fig.~\ref{Fig:swollowtail0} resemble the shape of a swallowtail. The swallowtail behavior for the CFT states demonstrates the existence of a first-order phase transition. When the central charge $C$ equals the critical value $C_{\text{c}}$, it becomes a second-order phase transition. However, the swallowtail disappears when the central charge $C$ is smaller than the critical value $C_{\text{c}}$.

As depicted in Fig.~\ref{swallow}, each swallowtail consists of three distinct piecewise smooth branches. The parameter $x$ increases along the curve, as evidenced by the direction of the arrows traced along the curve in Fig.~\ref{Fig:FTdiagram1}. Since the entropy of the CFT states is a monotonic increasing function of $x$, as indicated by eq.~(\ref{CFTS}), the entropy increases along the direction of the arrows depicted on the curve. The black line starting from zero temperature is the low entropy branch, since the value of $x$ is the smallest on this branch. The entropy increases along the read dashed curve, which is the intermediate entropy branch. The branch that the Helmholtz free energy $F$ extends to minus infinite is the largest entropy branch. For any swallowtail curve in the figure, starting at low temperature $\Tilde{T}$, the branch that minimizes the Helmholtz free energy is the thermodynamically favored state. Hence the low entropy branch is thermodynamically stable for low temperature. As the self-intersection point for the low entropy branch and high entropy branch has the same Helmholtz free energy, it is the phase transition point. The low and high entropy phases can exist together. After crossing the self-intersection point, the higher entropy branch is more stable. 

The stability of the CFT states can also be captured by the heat capacity. For positive heat capacity, the system is locally thermodynamically stable; while negative heat capacity indicates thermodynamically unstable. For the canonical ensemble with fixed $C, \mathcal{V}$ and $\tilde{\alpha}$, the heat capacity is
\begin{equation}
\begin{split}
  \mathcal{C}_{C, \mathcal{V}, \tilde{\alpha}}&=T\left(\frac{\partial S }{\partial \Tilde{T}}  
  \right)_{C, \mathcal{V}, \tilde{\alpha}}\\
  &= \frac{12 \pi  \text{C} \left(2 x^3+x\right) \left(x^2+2 y\right)^2}{x^2 (12 y-1)+2 x^4+2 y}.
  \end{split}
\end{equation}

\begin{figure}
  \centering
  \includegraphics[width=7cm]{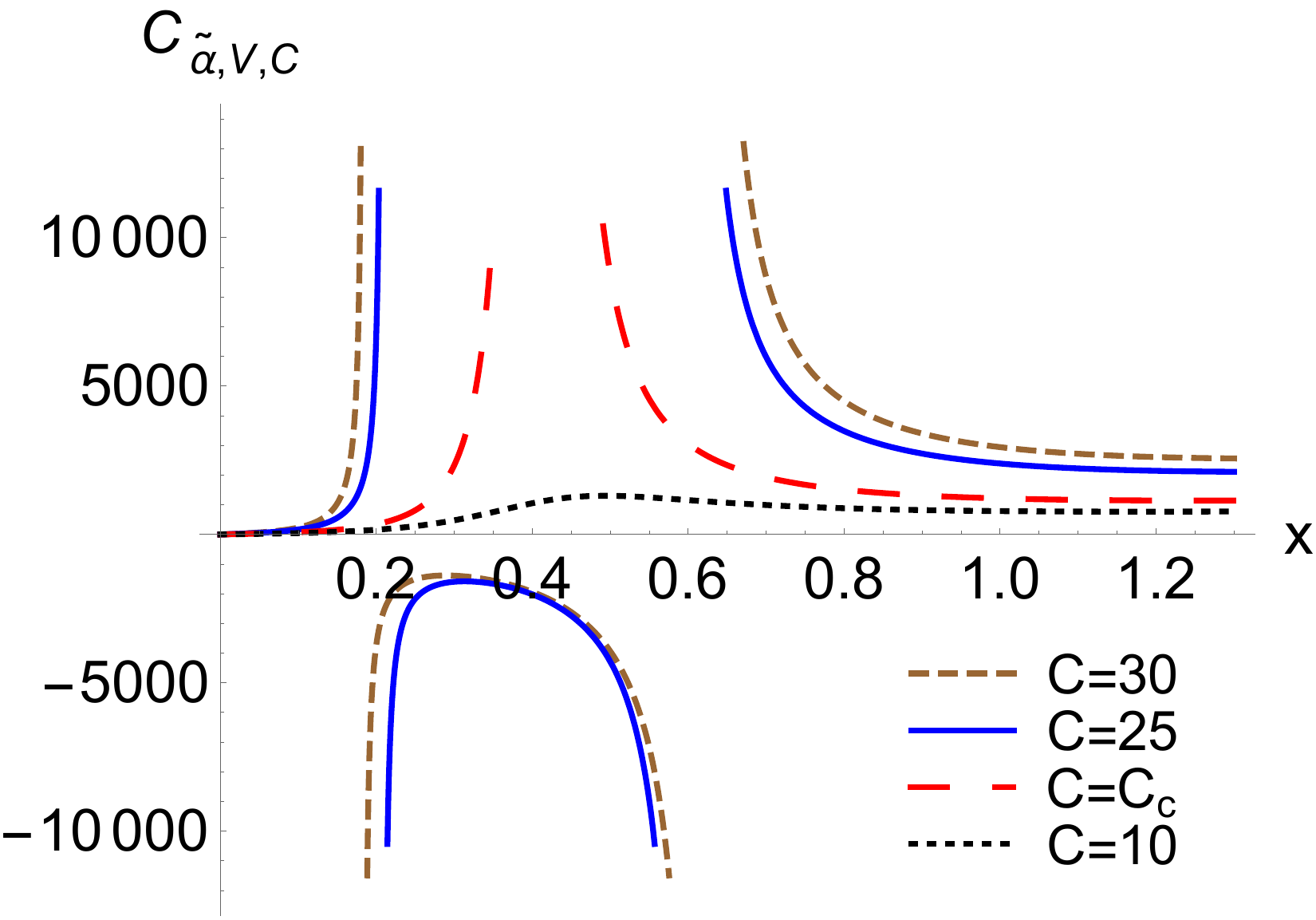}
  \caption{The heat capacity for the thermodynamic states of the boundary CFT, where we have set the Newtown's constant $G_{\text{N}}=1$, the boundary radius $R = 10,000$ and the CFT thermodynamic variable $\tilde{\alpha}=1$. For different curves the central charge are different. The heat capacity is continuous and finite for small central charge $C<C_{\text{c}}$; while for large central charge $C>C_{\text{c}}$, the heat capacity consists of three parts and there are points where the heat capacity is divergent.}\label{heatcapacity}
\end{figure}

The heat capacities for the low and high entropy branches of the Helmholtz free energy in Fig.~\ref{swallow} are both positive, hence they are thermodynamically stable. While the heat capacity for the intermediate entropy branch is negative, hence it is thermodynamically unstable. We also draw the heat capacity for the CFT states as in Fig.~\ref{heatcapacity}. For the brown dashed and blue line, the central charge $C$ is larger than the critical value $C_{\text{c}}=9\pi \tilde{\alpha}/\left(2\sqrt{G_{\text{N}}}\right) $. For the red large dashing line, the central charge is equal to the critical value $C = C_{\text{c}}$. For the black dotted line, the central charge $C$ is smaller than the critical value $C_{\text{}c}$. As we can see from the figure, the heat capacity is continuous and finite for small central charge $C< C_{\text{c}}$. However, there are points where the heat capacity is divergent for large central charge $C \geq  C_{\text{c}}$. The divergent points of the heat capacity correspond to the intersection points between the small/large entropy branches and intermediate entropy branches in Fig.~\ref{swallow}.

\begin{figure}
  \centering
  \includegraphics[width=7cm]{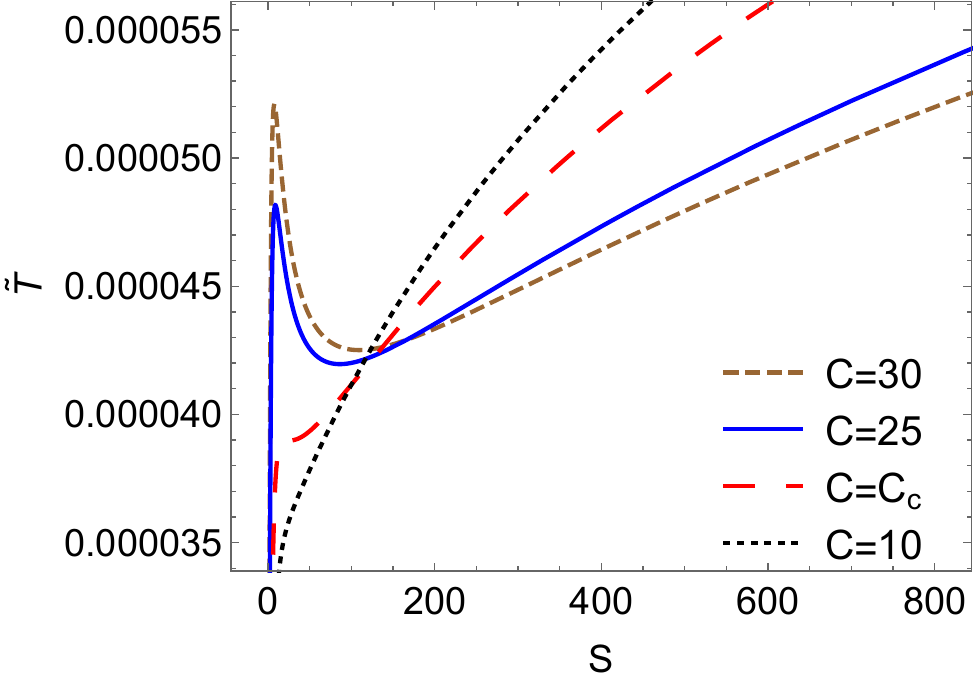}
  \caption{The $\tilde{T}-S$ oscillatory curves for the thermodynamics of the boundary CFT, where we have set the Newtown's constant $G_{\text{N}}=1$, the boundary radius $R = 10,000$ and the CFT thermodynamic variable $\tilde{\alpha}=1$. For different curves the central charge are different. For $C>C_{\text{c}}$, the brown dashed and blue curves in the $\tilde{T}-S$ plane display the oscillatory behavior. For $C\leq C_{\text{c}}$, the oscillatory behavior disappears, see the red dashed and black dotted lines. }\label{TSoscillatory0}
\end{figure}

Besides the Helmholtz free energy, the oscillatory behavior of the temperature $\Tilde{T}$ against the entropy $S$ can also depict the phase transition of the system. 
The oscillatory behavior for curves in the $\tilde{T}-S$ diagram in Fig.~\ref{TSoscillatory0} shows the phase transition of the boundary CFT. For large central charge $C>C_{\text{c}}$, the curves (the brown dashed and blue lines) in the $\tilde{T}-S$ plane display the oscillatory behavior. The oscillatory behavior of the $\tilde{T}-S$ diagram indicates there exists a first-order phase transition. For small central charge $C\leq C_{\text{c}}$, the oscillatory behavior disappears and the temperature $\tilde{T}$ is a monotonic increasing function of the entropy $S$, as can be seen in the figure for the red dashing and black dotted lines.

From the $\tilde{T}-S$ oscillatory curve in Fig.~\ref{TSoscillatory0}, we can get the critical point for the phase transition. The minimal and maximal points of the $\tilde{T}-S$ oscillatory curve coincide and exhibit a horizontal tangent at the critical point. Mathematically, the critical point satisfies the following equations:
\begin{equation}
    \begin{split}
        \left(\frac{\partial \tilde{T}}{\partial S}\right)_{C, \mathcal{V}, \tilde{\alpha}}=0, \qquad  \left(\frac{\partial^2\tilde{T}}{\partial S^2}\right)_{C, \mathcal{V}, \tilde{\alpha}}=0.
    \end{split}
\end{equation}
From the above two equations, we can get the critical point, which is determined by the following two parameters
\begin{equation}
    x_c=\frac{1}{\sqrt{6}}, \qquad y_c=\frac{1}{36}.
\end{equation}
Some straightforward calculations reveal that the critical point for the boundary CFT satisfies
\begin{equation}
    r_{\text{hc}}=\sqrt{6\alpha}.
\end{equation}
Evidently, the critical point of the thermodynamics for the boundary CFT is the same as the critical point for the bulk phase transition as indicated in eq.~\eqref{BHcritical}. 
At the critical point, the other thermodynamic quantities are
\begin{equation}
    \begin{split}
\tilde{T}_{\text{c}}&=\frac{\sqrt{6}}{2\pi R}, \qquad  \qquad 
\tilde{\mathcal{A}}_{\text{c}}=-\frac{9 \pi }{8\sqrt{G_{\text{N}}} R},  \\
 \mathcal{P}_{\text{c}}&= \frac{\tilde{\alpha}}{2 \pi\sqrt{G_{\text{N}} }R^4}, \qquad
S_{\text{c}}=\frac{\sqrt{6}}{\sqrt{G_{\text{N}}}}\pi^2\tilde{\alpha},  \\
C_{\text{c}}&=\frac{9\pi}{2\sqrt{G_{\text{N}}}} \tilde{\alpha},
\qquad \quad \mu_{\text{c}}=\frac{1}{4 R}.
    \end{split}
\end{equation}

From the critical point, we can easily find an interesting relation among the critical pressure $\mathcal{P}_\text{c}$, critical volume $\mathcal{V}_\text{c}$, critical temperature $\tilde{T}_\text{c}$, and the critical entropy $S_\text{c}$:
\begin{equation}
    \frac{\mathcal{P}_\text{c} \mathcal{V}_\text{c}}{\tilde{T}_\text{c} S_\text{c}}=\frac{1}{3}.
\end{equation}

\subsection{Maxwell's equal area law and coexistence curve}

The physical processes of crossing the coexistence phase often unveil intriguing phenomena, such as latent heat~\cite{Wei:2015iwa,Wei:2023mxw}. As shown by previous work~\cite{Zhou:2020vzf,Wei:2014qwa,Zhou:2019xai}, both the swallowtail behavior of the free energy and Maxwell's equal area law can reveal phase transition of a thermodynamic system and they are consistent with each other. In this subsection,  we derive Maxwell's equal area law from the change in the Helmholtz free energy during the phase transition between low and high entropy states. Then we employ Maxwell's equal area law to construct an analytical formula for the coexistence curve in the $\tilde{T}-S$ plane.

To solve the coexistence curve for the low and high entropy states of the boundary CFT, we need to solve the phase transition points in the $F-\tilde{T}$ plane in Fig.~\ref{swallow}. We focus on the self-intersection points in Fig.~\ref{Fig:FTdiagram1}. These two points are low and high entropy phases. These two coexistence phases correspond to point A and point C in the $\Tilde{T}-S$ plane, respectively. A phase of the boundary CFT thermodynamic system in the fixed $(C, \mathcal{V}, \tilde{\alpha})$ canonical ensemble can be determined by the parameter $x$. We denote the low entropy phase A as $ (x_{\text{L}},y,R,C)$ and the high entropy phase C as $(x_{\text{H}},y,R,C)$.

\begin{figure}
    \centering
    \includegraphics[width=8cm]{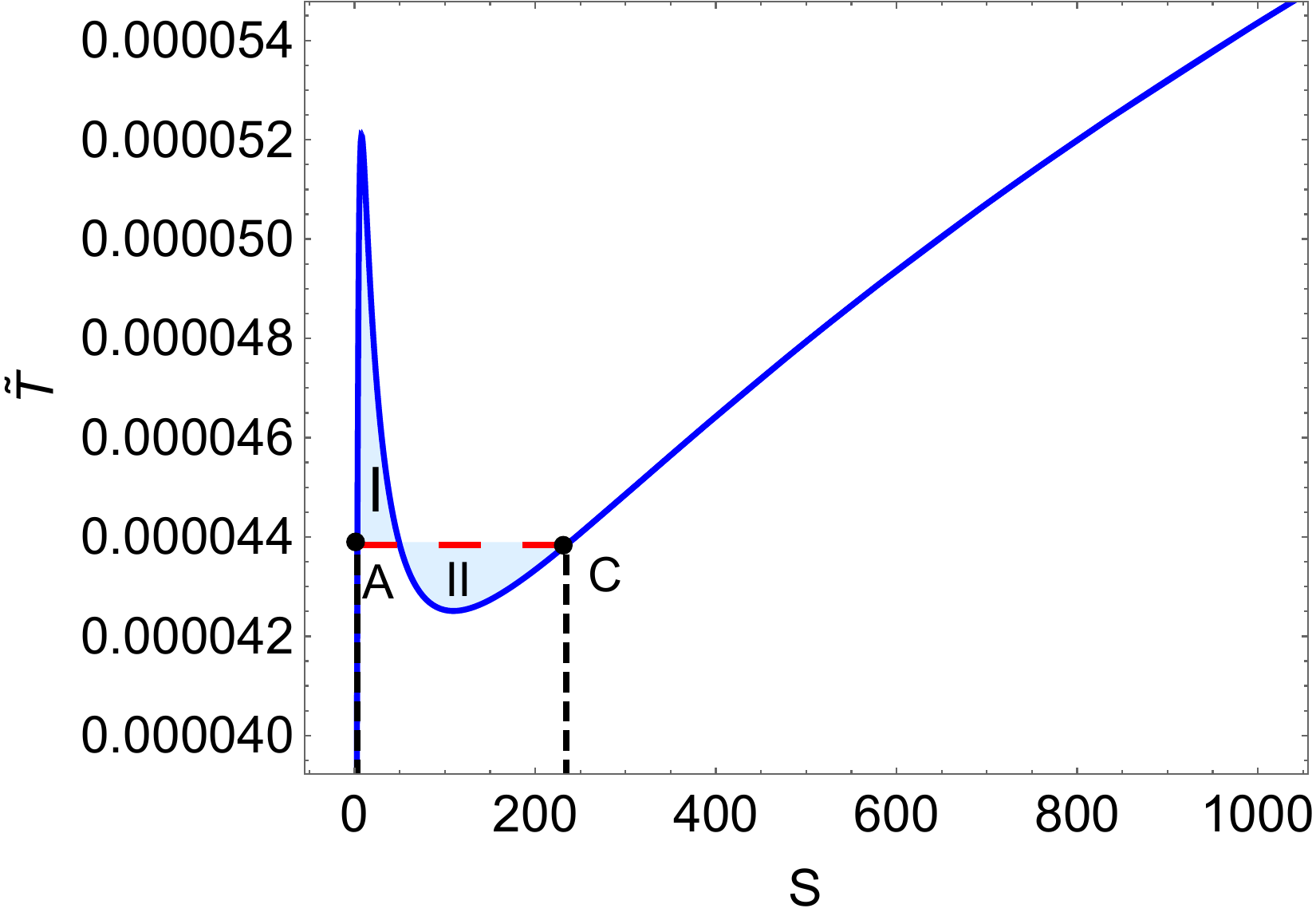}
    \caption{Maxwell's equal area law: For large central charge $C>C_{\text{c}}$, the temperature $\Tilde{T}$ vs the entropy $S$ exhibits oscillatory behavior, which indicates a first-order phase transition. At the phase transition temperature the areas of the two shadowed regions are the same. We have set the Newtown's constant $G_{\text{N}}=1$, the boundary radius $R = 10,000$, the central charge $C=30$, and the CFT thermodynamic variable $\tilde{\alpha}=1$ in drawing the figure. }
    \label{fig:MaxwellEqAl}
\end{figure}

For the phase transition between the low entropy state A and high entropy state C in Fig.~\ref{Fig:FTdiagram1}, the Helmholtz free energies of the two phases are the same and the difference is zero:
\begin{equation}
    \Delta F=0.\label{DeltaF}
\end{equation}
 From the Helmholtz free energy
\begin{equation}\label{CaF}
    F=\tilde{E}-\tilde{T}S,
\end{equation}
and the first law~\eqref{1stlaw}, we have
\begin{equation}
    dF=Sd\tilde{T}-\mathcal{P}d\mathcal{V}+\Tilde{\mathcal{A}}d\Tilde{\alpha}+\mu dC.
\end{equation}
Integrating it and referring to eq.~\eqref{DeltaF}, we obtain
\begin{equation}
\begin{split}
    \int^{\text{C}}_\text{A} dF&=\int^{\text{C}}_\text{A} Sd\Tilde{T}-\int^{\text{C}}_\text{A} Pd\mathcal{ V}+\int^{\text{C}}_\text{A} \tilde{\mathcal{A}}d\Tilde{\alpha}+\int^{\text{C}}_\text{A}\mu dC\\
    &=0.\label{interTS}
    \end{split}
\end{equation}
Evidently, the exactly result of each integral in eq.~\eqref{interTS} depends on the physical process.
For the fixed $C$, $\mathcal{V}$, and $\Tilde{\alpha}$ canonical ensemble, eq.~\eqref{interTS} becomes
\begin{equation}
\int^{\text{C}}_\text{A} Sd\Tilde{T}=0.\label{MEAL}
\end{equation}
The result shows that the areas in the region $\uppercase\expandafter{\romannumeral1}$ and region $\uppercase\expandafter{\romannumeral2}$ in Fig.~\ref{fig:MaxwellEqAl} are the same. It suggests that Maxwell's equal area law holds for the phase transition of the boundary CFT in the canonical ensemble. As we can see from Fig.~\ref{fig:MaxwellEqAl},
it is more convenient and direct to formulate Maxwell's equal area law~\eqref{MEAL} in the following form
\begin{equation}
    \int^{\text{C}}_\text{A} \Tilde{T}dS=\Tilde{T}(\text{A})\left[S(\text{C})-S(\text{A})\right].\label{MaxwellA}
\end{equation}
From the temperature in eq.~\eqref{CFTT} and the entropy in eq.~\eqref{CFTS} of the CFT, we can get the integration, which is
\begin{equation}
    \begin{split}
        \int^{\text{C}}_\text{A} \Tilde{T}dS&=\int^{x_{\text{H}}}_{x_{\text{L}}}\Tilde{T}\left(\frac{\partial S}{\partial x}\right)dx\\
        &=\frac{3C}{R}(x_{\text{L}}^2+x_{\text{H}}^2+1)(x_{\text{L}}+x_{\text{H}})(x_{\text{H}}-x_{\text{L}}).
    \end{split}
\end{equation}
Besides, the right-hand side of eq.~\eqref{MaxwellA} is
\begin{multline}
           \tilde{T}(\text{A})\left[S(\text{C})-S(\text{A})\right]
       =\frac{2C}{R}\frac{2x_{\text{L}}^3+x_{\text{L}}}{x_{\text{L}}^2+2y}\\
       \times(x_{\text{L}}^2+x_{\text{L}}x_{\text{H}}+x_{\text{H}}^2+6y)(x_{\text{H}}-x_{\text{L}}).
\end{multline}
Then, Maxwell's equal area law~\eqref{MaxwellA} indicates the following equation
\begin{multline}
    3(x_{\text{L}}+x_{\text{H}})(x_{\text{L}}^2+2y)(x_{\text{L}}^2+x_{\text{H}}^2+1)
        =2(2x_{\text{L}}^3+x_{\text{L}})\\
        \times(x_{\text{L}}^2+x_{\text{L}}x_{\text{H}}+x_{\text{H}}^2+6y).
\end{multline}
On the other hand, the temperatures of the phase A and phase C are the same:
\begin{equation}
    \tilde{T}(\text{A})= \tilde{T}(\text{C}). \label{equalT}
\end{equation}

\begin{figure}
    \centering
    \includegraphics[width=8cm]{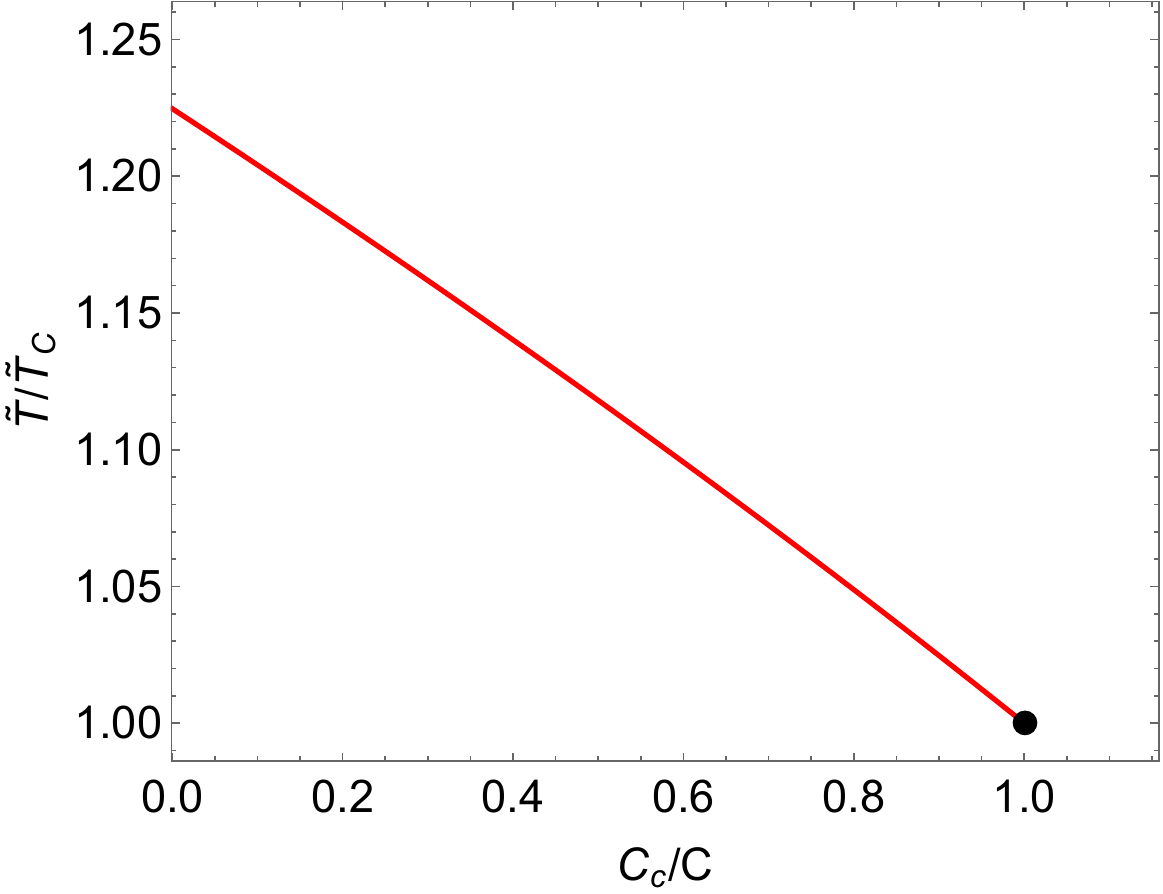}
    \caption{The coexistence curve: The temperature decreases with the inverse of the central charge $C$. The solid black dot stands for the critical point. For $C> C_{\text{c}}$, a first-order phase transition occurs between the low and high entropy states. In the critical point $C=C_{\text{c}}$, a second-order phase transition occurs. }
    \label{fig:coexistencecurve}
\end{figure}

From Maxwell's equal area law in eq.~\eqref{MaxwellA} and eq.~\eqref{equalT}, we can get the parameters characterizing the low entropy phase A and high entropy phase C:
\begin{equation}
    \begin{split}
        x_{\text{L}}&=\frac{\sqrt{1-24 y-\sqrt{48 y (9 y-1)+1}}}{\sqrt{2}}, \\
        x_{\text{H}}&=\frac{6 \sqrt{2} y}{\sqrt{1-24 y-\sqrt{48 y (9 y-1)+1}}}.\label{LHposition}
    \end{split}
\end{equation}
The isothermal temperature for phases A and C, representing the temperature of coexistence, can be determined. The temperature indicates the precise conditions at which coexistence takes place. The isothermal temperature is
\begin{multline}
    \tilde{T}=\frac{\sqrt{1-24 y-\sqrt{48 y (9 y-1)+1}}}{\sqrt{2} \pi  R \left(1-20 y-\sqrt{48 y (9 y-1)+1}\right)}\\
    \times  \left(2-24 y-\sqrt{48 y (9 y-1)+1}\right).\label{transitionT}
\end{multline}
Since we are considering the fixed $(C, \mathcal{V}, \tilde{\alpha})$ canonical ensemble, the function for the coexistence curve for the canonical ensemble is given by eq.~\eqref{transitionT} and
\begin{equation}
    \begin{split}
        C=\frac{\pi}{8\sqrt{G_{\text{N}}}}\frac{\tilde{\alpha}}{y}.\label{EQ:coexistencecurve}
\end{split}
\end{equation}
The above equations~\eqref{transitionT} and~\eqref{EQ:coexistencecurve} are the parameter equations for the coexistence curve. The temperature for the coexistence of the two phases is a function of the parameter $y$ and boundary volume $\mathcal{V}$. To write the equation in a more concise manner, we define
\begin{equation}
    \begin{split}
        \tau=\frac{\tilde{T}}{\tilde{T}_{\text{c}}}, \qquad c=\frac{C}{C_{\text{c}}}.\label{dimensionlessP}
    \end{split}
\end{equation}
The coexistence curve equation can be written as
\begin{multline}
    \tau=\sqrt{-\sqrt{3} \sqrt{\frac{1}{c^2}-\frac{4}{c}+3}-\frac{2}{c}+3}\\
    \times\frac{c \left(\sqrt{3} \sqrt{\frac{1}{c^2}-\frac{4}{c}+3}-6\right)+2}{c \left(3 \sqrt{3} \sqrt{\frac{1}{c^2}-\frac{4}{c}+3}-9\right)+5}.
\end{multline}
One of the most remarkable properties of the above equation for the coexistence curve in the reduced parameter space is its independence of the CFT thermodynamic variable $\tilde{\alpha}$, as well as the boundary volume $\mathcal{V}$. This behavior is analogous to the phase transition in the $5$D neutral Gauss-Bonnet AdS black hole, where the coexistence curve is independent of the Gauss-Bonnet coupling constant $\alpha $ when expressed in the reduced parameter space, as discussed in~\cite{Mo:2015scl}. From the coexistence curve in Fig.~\ref{fig:coexistencecurve}, the temperature for the low and high entropy coexistence phases decreases with the increasing of the inverse of the central charge $1/C$.
In the coexistence curve, a first-order phase transition occurs between the low and high entropy states for $C> C_{\text{c}}$. At the critical point, a second-order phase transition occurs. However, there is no phase transition for $C<C_{\text{c}}$.

\subsection{Critical exponents}

Physics near the critical point often exhibits fascinating phenomena. Critical exponents characterize the behavior of physical quantities in the vicinity of a critical point. These critical exponents are considered universal, as they remain independent of the specific characteristics of the physical system. In this subsection, we study the critical behavior of some physical quantities and calculate the corresponding critical exponents for the thermodynamics of the CFT residing on the boundary.

In the vicinity of the critical point, the critical behavior of the CFT dual to the five-dimensional neutral Gauss-Bonnet AdS black hole can be precisely characterized by the critical exponents denoted as $\alpha', \beta, \gamma$, and $\delta$. These critical exponents are explicitly defined as follows:
\begin{itemize}
\item The critical exponent $\alpha'$ governs the behavior of the heat capacity of the CFT at constant chemical potential as compared to the heat capacity of an AdS black hole at constant volume
\begin{equation}
\begin{split}
         C_{\mu}=\tilde{T}\left(\frac{\partial S}{\partial\tilde{T}}\right)_{\mu}\sim\left(-\frac{\Tilde{T}-\Tilde{T}_{\text{c}}}{\Tilde{T}_{\text{c}}}\right)^{-\alpha'}.
    \end{split}
\end{equation}
\item  The critical exponent $\beta$ determines the behavior of the order parameter $\eta$. Since the entropy increases along the arrows in the swallowtail curve in Fig.~\ref{Fig:FTdiagram1}, and the difference of the entropies $S_{\text{H}}-S_{\text{L}}$ changes discontinuously in the first-order phase transition but continuously at the critical point, we choose the entropy difference between high and low entropy phases as the order parameter. Then the critical exponent $\beta$ is defined as
\begin{equation}
    \begin{split}
     \eta=  \frac{S_{\text{H}}-S_{\text{L}}}{S_{\text{c}}}\sim  \left(-\frac{\tilde{T}-\Tilde{T}_{\text{c}}}{\Tilde{T}_{\text{c}}}\right)^{\beta}.
    \end{split}
\end{equation}
\item The critical exponent $\delta $ determines the behavior of the pressure on the critical point for extended black hole thermodynamics. Through the dual relation, the critical exponent $\delta$ for the CFT is defined as
\begin{equation}
    \begin{split}
    C-C_{\text{c}} \sim (  \mu-\mu_{\text{c}})^\delta.
    \end{split}
\end{equation}
\end{itemize}

These critical exponents characterize the critical phenomena for the boundary CFT. To calculate the critical exponents for the CFT, we define
\begin{equation}
    t=\frac{\tilde{T}}{\tilde{T}_{\text{c}}}-1, \qquad
    s=\frac{S}{S_{\text{c}}}-1, \qquad \chi=\frac{\mu}{\mu_{\text{c}}}-1.
\end{equation}
At the critical point, the heat capacity for the fixed chemical potential is
\begin{equation}
    C_{\mu,\mathcal{V},\tilde{\alpha}}=\tilde{T}\left(\frac{\partial S}{\partial\tilde{T}}\right)_{\mu,\mathcal{V},\tilde{\alpha}}=\frac{4 \sqrt{6}  }{\sqrt{G\text{N}}}\pi^2 \tilde{\alpha},
\end{equation}
which is not divergent.  Hence the critical exponent  governing the behavior of the heat capacity for the CFT is $\alpha'=0$.

To calculate the critical exponent governing the behavior of the order parameter, we start at the entropy of the CFT at the phase transition point. Together with the entropy~\eqref{CFTS}, and the parameter labeling the low and high entropy states~\eqref{LHposition}, the entropy of the phases A and C in Fig.~\ref{fig:MaxwellEqAl} are
\begin{align}
    S_{\text{L}}&=\sqrt{2} \pi  C \sqrt{1-24 y-\sqrt{48 y (9 y-1)+1}} \notag \\
    &\times\left(1-12 y-\sqrt{48 y (9 y-1)+1}\right),\\
    S_{\text{H}}&=\frac{144 \sqrt{2} \pi  C y^2 \left(1-12 y-\sqrt{48 y (9 y-1)+1}\right)}{\left(1-24 y-\sqrt{48 y (9 y-1)+1}\right)^{3/2}}.
\end{align}
In terms of the parameter $y$, the order parameter can be expressed as
\begin{equation}
\begin{split}
    \eta&=\frac{S_{\text{H}}-S_{\text{L}}}{S_{\text{c}}}\\
    &=\left(12 y+\sqrt{48 y (9 y-1)+1}-1\right)^2 \\
    &\times\frac{ 36 y+\sqrt{48 y (9 y-1)+1}-1}{8 \sqrt{3} y \left(-24 y-\sqrt{48 y (9 y-1)+1}+1\right)^{3/2}}.\label{oederPara}
\end{split}
\end{equation}
From the temperature $\tilde{T}$ in eq.~\eqref{transitionT},  we have
\begin{equation}
\begin{split}
     t&=\frac{\tilde{T}}{\tilde{T}_{\text{c}}}-1\\
     &=\sqrt{1-24 y-\sqrt{48 y (9 y-1)+1}}\\
     &\times \frac{ 24 y+\sqrt{48 y (9 y-1)+1}-2}{\sqrt{3} \left(20 y+\sqrt{48 y (9 y-1)+1}-1\right)}-1.\label{Temperay}
\end{split}
\end{equation}
Solving eq.~\eqref{Temperay}, we can get a simple result for the parameter $y$ in terms of $t$:
\begin{equation}
    y= \frac{1}{36} \left(-2 t^2-4 t+1\right).\label{yt}
\end{equation}
Together with eqs.~\eqref{oederPara} and~\eqref{yt}, the order parameter can be formulated in terms of the temperature parameter $t$:
\begin{equation}
    \begin{split}
        \eta &=\frac{4 \left(t (t+2)-\sqrt{3} \sqrt{t (t+1)^2 (t+2)}+1\right)^2}{1-2 t (t+2)}\\
        &\times\frac{ \sqrt{3} \sqrt{t (t+1)^2 (t+2)}-3 t (t+2)}{ \left(4 t (t+2)-2 \sqrt{3} \sqrt{t (t+1)^2 (t+2)}+1\right)^{3/2}}\\
        &=4\sqrt{6}t^{\frac{1}{2}}+25\sqrt{6}t^{\frac{3}{2}}+\mathcal{O}(t^{\frac{5}{2}}).
    \end{split}
\end{equation}
Hence, the critical exponent is given by $\beta=1/2$.

To calculate the critical exponent $\delta$, through the definition of $t$ and the equation for the temperature~\eqref{CFTT}, we can solve the third-order equation for $x$ in terms of the parameters $t$ and $y$. Then the chemical potential $\chi$ can be formulated as a function of the parameters $t$ and $y$. 
Taking $y=y_{\text{c}}=1/36$ and expand $\chi$ near the critical point, we obtain
\begin{equation}
    \begin{split}
        \chi\approx\frac{(1+3c)(c-1)^\frac{1}{3}}{3c^{\frac{4}{3}}}\sim \left(c-1\right)^\frac{1}{3},
    \end{split}
\end{equation}
where $c$ is the reduced central charge as defined in eq.~\eqref{dimensionlessP}. Then we have
\begin{equation}
    c-1 \sim \chi^3=\left(\frac{\mu}{\mu_{\text{c}}}-1\right)^3.
\end{equation}
Hence, the critical exponent is $\delta=3$.

Thus, we have obtained the critical exponents $\alpha'$, $\beta$ and $\delta$ associated with the $\tilde{T}-S$ criticality of the CFT,  where the conformal factor $\omega$ is free to vary but with fixed Newton's constant. Our results suggest that these critical exponents associated with the $\tilde{T}-S$ criticality of the CFT are the same as those of the five-dimensional Gauss-Bonnet AdS black hole~\cite{CCLY13,Miao:2018fke}.

\section{Conclusion and discussion}\label{4section}

In the AdS/CFT correspondence, black holes in AdS spacetime correspond to thermal states of the boundary field theories. Based on the recently proposed viewpoint of the holographic thermodynamics, which treats the conformal factor $\omega$ as a variable, we derived the first law and Euler equation for the thermodynamics of the CFT, and established a holographic dictionary between the thermodynamics of the five-dimensional  neutral Gauss-Bonnet AdS black hole and the dual boundary thermodynamics. Having the thermodynamics for the dual CFT, we investigated the phase transition and criticality of the CFT in the fixed $(C, \mathcal{V}, \tilde{\alpha})$  canonical ensemble. Just like the free energy for the five-dimensional neutral Gauss-Bonnet AdS black hole, the free energy for the CFT displays a characteristic swallowtail behavior. This indicates a first-order phase transition below the critical point and a second-order phase transition at the critical point. Through the $\tilde{T}-S$ oscillatory behavior, we obtained the critical point for the thermodynamics of the CFT in the  $\tilde{T}-S$ plane. The critical point for the CFT is exactly the same as that of the five-dimensional neutral Gauss-Bonnet AdS black hole. Using Maxwell's equal area law, we got the coexistence curve for the high and low entropy phases of the CFT. Besides, we got the critical exponents for the CFT, and found that these critical exponents associated with the $\tilde{T}-S$ criticality are the same as those of the five-dimensional Gauss-Bonnet AdS black hole. We have chosen the CFT thermodynamical variable $\tilde{\alpha}=1$ in drawing some of our figures, but the qualitative characteristics remain unaffected by the specific values we select.

Critical exponents describe the singularity behavior of thermodynamic variables such as the heat capacity, order parameter, etc. They are universal and satisfy the scaling hypothesis. The critical exponent $\gamma$ governing the isothermal compressibility of black hole thermodynamics is defined as
\begin{equation}
    \begin{split}
     \kappa_{T}=-\frac{1}{v}\left(\frac{\partial v}{\partial P}\right)_{T}\sim \left(-\frac{T-T_{\text{c}}}{T_{\text{c}}}\right)^{-\gamma}.
    \end{split}
\end{equation}
As we argued in Sec.~\ref{boundarythermo}, the black hole thermodynamic pressure $P$ is dual to the CFT central charge $C$, and the black hole thermodynamic volume $V$ is dual to the CFT chemical potential $\mu$.
In our investigation of the critical exponents for the CFT thermodynamics in the $\tilde{T}-S$ plane, we found that if we define the critical exponent $\gamma$ just by replacing the pressure $P$ and the  specific volume $v$ by the central charge $C$ and the chemical potential $\mu$ respectively
\begin{equation}
    \begin{split}
     \kappa_{\tilde{T}}=-\frac{1}{\mu}\left(\frac{\partial \mu}{\partial C}\right)_{\tilde{T},\mathcal{V},\tilde{\alpha}}\sim \left(-\frac{\Tilde{T}-\Tilde{T}_{\text{c}}}{\Tilde{T}_{\text{c}}}\right)^{-\gamma},\label{gamma}
    \end{split}
\end{equation}
then, the critical exponent is given by $\gamma=2/3$, which is different from the prediction of the scaling hypothesis. Perhaps this is due to the fact that it is the black hole thermodynamic volume $V$ instead of the specific volume $v$ that is dual to the CFT chemical potential $\mu$. This leads to the result that the isothermal compressibility is not defined just by replacing the pressure $P$ and the  specific volume $v$ by the central charge $C$ and the chemical potential $\mu$ in the $\tilde{T}-S$ plane.

Criticality is a very important phenomenon for thermodynamic systems, and critical exponents are universal. Our investigation of the thermodynamics and criticality from the gravity/gauge duality viewpoint might lead to new insight into the understanding of the Van der Waals like phase behavior of the five-dimensional Gauss-Bonnet AdS black hole.

We conclude with a final remark on the construction of CFT thermodynamics from bulk thermodynamics for higher-dimensional black holes and modified gravity theories. The number of central charges depends on the spacetime dimension. For higher-dimensional black holes in AdS space, two or more central charges may arise. Additionally, the value of the central charge can vary for different gravitational theories and they might not be the same value. However, the construction method proposed in Ref.~\cite{ACKMV23a} yields a single central charge. This observation is worth further investigation.

\acknowledgments

We are grateful to Yi Pang, Hong L\"u, Chuan Chen, Shan-Ping Wu, and Yu-Peng Zhang for useful discussions. We thank the Center for Joint Quantum Studies and Department of Physics (Tianjin University), where part of this work has
been done, for its hospitality. This work was supported by National Key Research and Development Program of China (Grant No. 2021YFC2203003), the National Natural Science Foundation of China (Grants No. 12305065, No. 12247178, No. 12475056, No. 12347177,  No. 12075103 and No. 12247101), the China Postdoctoral Science Foundation (Grant No. 2023M731468), Gansu Province's Top Leading Talent Support Plan, and the 111 Project (Grant No. B20063).\\

%

\end{document}